\documentclass{webofc}
\usepackage[varg]{txfonts}   
\begin{document}
\title{Near-threshold hadron scattering with effective field theory}
%
%

\author{\firstname{Katsuyoshi} \lastname{Sone}\inst{1}\fnsep\thanks{\email{sone-katsuyoshi@ed.tmu.ac.jp}} \and
        \firstname{Tetsuo} \lastname{Hyodo}\inst{1}\fnsep\thanks{\email{hyodo@tmu.ac.jp}} 
}

\institute{Department of Physics, Tokyo Metropolitan University, Hachioji 192-0397, Japan
          }

\abstract{%
When an exotic hadron locates near the threshold with the channel couplings, the internal structure of the exotic hadron is related to the scattering length. To incorporate the threshold effect, the Flatt\'{e} amplitude has been often used to determine the scattering length. It is however known that an additional constraint is imposed on the Flatte amplitude near the threshold. We discuss this problem by using the effective field theory for the coupled-channel scattering.
}
\maketitle
\section{Introduction}
\label{intro}

Exotic hadrons, such as $T_{cc}$, $X(3872)$, and $f_{0}(980)$, are currently attracting attention~\cite{Guo:2017jvc,Brambilla:2019esw}. Many exotic hadrons are known to appear near the threshold of two hadron scattering. In such cases, the internal structure of exotic hadrons is strongly related to the scattering length. When there is not only one scattering channel but also decay channels, it is also necessary to consider the effect of the channel couplings. For the analysis of near-threshold exotic hadrons, the Flatt\'{e} amplitude~\cite{Flatte:1976xu} is now often used, which includes the threshold effect. Since each component of the Flatt\'{e} amplitude can be written in the form of the effective range expansion, the scattering length $a_{\rm F }$ can be determined from the Flatt\'{e} amplitude.

However, the Flatt\'{e} amplitude has the following problem; in the case of the two-channel scattering, the Flatt\'{e} amplitude has three parameters, but the number of parameters is reduced to two near the threshold~\cite{Baru:2004xg}. Thus, the Flatt\'{e} amplitude is not a general amplitude in the threshold region, and the Flatt\'{e} scattering length $a_{\rm F }$ may not be general. In more general frameworks, how would the scattering length be described?

\section{Comparison of Flatt\'{e} and EFT}
\label{sec-1}
We take the case of the two-channel scattering as an example and compare the Flatt\'{e} amplitude with the general form of the scattering amplitude determined from the optical theorem. We consider the case that the threshold of channel 2 is higher than that of channel 1. The Flatt\'{e} amplitude at energy $E$ is written by
\begin{align}
     f^{\rm F} = \frac{1}{-2E+2E_{\rm BW} - ig_{1}^2p - ig_2^2k} 
     \begin{pmatrix}
         g_1^2 & g_1g_2 \\
         g_1g_2 & g_2^2
     \end{pmatrix}, \label{eq: Flatte amplitude}
\end{align}
where $g_{1}$ and $\ g_{2}$ represent the coupling contants and $E_{\rm BW}$ is the bare energy. $p(E)$ and $k(E)$ denote the momenta in channels 1 and 2, respectively. It is known that the Flatt\'{e} amplitude satisfies the optical theorem. However, when $f^{\rm F}$ is expanded up to the first order in $k$ near the channel 2 threshold, it is known that the amplitude depends only on $R=g_2^2/g_1^2$ and $\alpha=2E_{\rm BW}/(g_1^2p_0)$, reducing the number of independent parameters to two~\cite{Baru:2004xg}, where $p_0=p(E=0)$ is the momentum of channel 1 at the threshold of channel 2.

On the other hand, K-matrix, M-matrix, etc. are known as general scattering amplitudes satisfying the optical theorem~\cite{Badalian:1981xj}. In this study, we use the EFT amplitude~\cite{Cohen:2004kf} derived from the non-relativistic effective field theory (EFT) with contact interactions:
\begin{align}
     f^{\rm EFT} = \frac{1}{\frac{1}{a_{12}^2} - \left(\frac{1}{a_{11}} + ip\right)\left(\frac{1}{a_{22}} + ik\right)}
     \begin{pmatrix}
         \frac{1}{a_{22}} + ik & \frac{1}{a_{12}} \\
         \frac{1}{a_{12}} & \frac{1}{a_{11}} + ip
     \end{pmatrix}.
     \label{eq:EFT amplitude}
 \end{align}
$a_{11},a_{12},a_{22}$ are the parameters of the EFT amplitude in units of the length. The EFT amplitude, like the Flatt\'{e} amplitude, contains three parameters and satisfies the optical theorem. We expand the EFT amplitude up to the first order in $k$ and it is shown that the number of parameters of the EFT amplitude does not decrease and remains three even near the threshold. Therefore, the use of the EFT amplitude solves the problem of the Flatt\'{e} amplitude.

Although the EFT amplitude is found to be more general than the Flatt\'{e} amplitude, the relationship between the EFT amplitude and the Flatt\'{e} amplitude is not clear. This is because the EFT amplitude has an inverse matrix, while the Flatt\'{e} amplitude does not, and thus the EFT amplitude and the Flatt\'{e} amplitude cannot be directly mapped to each other. In order to clarify the relationship between the two, we construct a scattering amplitude that can represent both the EFT amplitude and the Flatt\'{e} amplitude.

\section{General amplitude}
\label{sec-2}
We introduce the general amplitude $f^{\rm G}$ with a new parametrization based on the EFT amplitude. $f^{\rm G}$ is represented by dimensionless constants $\gamma$ and $\epsilon$ and a parameter $A_{22}$ in units of the length:
\begin{align}
     f^{\rm G} &= \frac{1}{-\frac{1}{A_{22}^2} - i\frac{1}{A_{22}}\epsilon p - i\frac{1}{A_{22}}k - \gamma pk}
     \begin{pmatrix}
         \frac{1}{A_{22}}\epsilon + i\gamma k & \frac{1}{A_{22}}\sqrt{\epsilon-\gamma} \\
         \frac{1}{A_{22}}\sqrt{\epsilon-\gamma} &  \frac{1}{A_{22}} + i\gamma p
     \end{pmatrix}. 
\end{align}
When $\gamma = \epsilon$, the channel couplings vanish and $f^{\rm G}$ is written by $f^{\rm G}=(1/(-1/(A_{22}\epsilon)-ip),1/(-1/(A_{22})-ik))$. From this, $A_{22}$ represents the scattering length of channel 2 in the absence of the channel couplings.

Next, we discuss the relation between the general amplitude, the EFT amplitude, and the Flatt\'{e} amplitude. When $\gamma=0$, $f^{\rm G}$ is given as follows:
\begin{align}
     f^{\rm G} &= \frac{1}{-\frac{1}{A_{22}} - i\epsilon p_0 -ik}
     \begin{pmatrix}
          \epsilon && \sqrt{\epsilon} \\
          \sqrt{\epsilon} && 1
     \end{pmatrix}.
\end{align}
This amplitude is equivalent to the Flatt\'{e} amplitude up to first order in $k$. In other words, the general amplitude with $\gamma=0$ reproduces the Flatt\'{e} amplitude. Furthermore, the decreas of the Flatt\'{e} amplitude parameters near the threshold can be understood from the condition $\gamma=0$ in the general amplitude. In this case, the determinant of the general amplitude behaves as $\lim_{\gamma \to 0}  {\rm det}\left(f^{\rm G}\right) = 0$. From this feature, the inverse of $f^{\rm G}$ does not exist when $\gamma=0$. Since general amplitude is obtained by a different parametrization of the EFT amplitude, for $\gamma \neq 0$, the general amplitude corresponds to the EFT amplitude. In other words, both the Flatt\'{e} amplitude and the EFT amplitude can be obtained from the general amplitude by choosing the parameter $\gamma$.

Next, we perform the effective range expansion for $f^{\rm G}$ in terms of $k$, and determine the scattering length. We expand the denominator of $f^{\rm G}_{22}$ in powers of the momentum $k$:
\begin{align}
     f^{\rm G}_{22} = \frac{1}{-\frac{1}{A_{22}}\left(\frac{\frac{1}{A_{22}} + i\epsilon p_0}{\frac{1}{A_{22}} + i\gamma p_0}\right) - \frac{i\left(\epsilon - \gamma\right)}{2\left(1+iA_{22}\gamma p_0\right)p_0^2}k^2 - ik + O(k^4)}. \label{eq: ERE of fG22}
\end{align}
This shows that, $f_{22}^{\rm G}$ can be written as the effective range expansion in $k$, and we can define the scatterng length $a_{\rm G}$ in the general amplitude as follows:
\begin{align}
     a_{\rm G} = A_{22} \left(\frac{\frac{1}{A_{22}} + i\gamma p_0}{\frac{1}{A_{22}} + i\epsilon p_0}\right). \label{eq: the scattering length of fG}
\end{align}
In the same way, we expand $f^{\rm G}_{11}$ in $k$:
\begin{align}
     f^{\rm G}_{11} = \frac{\frac{\epsilon^2}{\epsilon - \gamma}}{-\frac{1}{A_{22}}\frac{\epsilon}{\epsilon - \gamma} - i\frac{\epsilon^2}{\epsilon - \gamma}p_0 - \left(A_{22}\frac{\gamma}{\epsilon} + i\frac{\epsilon^2}{2\left(\epsilon - \gamma\right)p_0}\right) - ik + O(k^3)}. \label{eq: expansion of fG11 in terms of k}
\end{align}
Because the power series in Eq.~\eqref{eq: expansion of fG11 in terms of k} contains terms such as $k^3$, $f^{\rm G}_{11}$ cannot be written in the form of the effective range expansion. Also, the coefficients of each term in the denominator of $f_{11}^{\rm G}$ are different from those of $f_{22}^{\rm G}$ in Eq.~\eqref{eq: ERE of fG22}. In particular, the constant term in the denominator of $f^{\rm G}_{11}$ is different from the scattering length in Eq.~\eqref{eq: the scattering length of fG}. On the other hand, as can be seen from Eq.~\eqref{eq: Flatte amplitude}, the coefficients of the power series of the Flatt\'{e} amplitude are common for all the components, and the constant terms of the denominator of the scattering amplitudes are entirely given by the Flatt\'{e} scattering length.

To summarize, from Eqs.~\eqref{eq: ERE of fG22} and \eqref{eq: expansion of fG11 in terms of k}, in general, the $f_{22}$ component can be written as the effective range expansion near the threshold of channel 2, but the $f_{11}$ component cannot be written and the scattering length cannot be defined. On the other hand, when $\gamma = 0$, $f^{\rm G}$ reduces to the Flatt\'{e} amplitude, and the scattering length $a_{\rm G}$ becomes the Flatt\'{e} scattering length $a_{\rm F}$ as follows:
\begin{align}
     a_{\rm G} = A_{22} \left(\frac{\frac{1}{A_{22}} + i\gamma p_0}{\frac{1}{A_{22}} + i\epsilon p_0}\right) \xrightarrow{\gamma=0} \frac{1}{\frac{1}{A_{22}} + i\epsilon p_0}=a_{\rm F} . \label{eq: the limit gamma to 0}
\end{align}
Similarly, the constant term of the denominator of $f_{11}^{\rm G}$ in Eq.~\eqref{eq: expansion of fG11 in terms of k} becomes $a_{\rm F}$:
\begin{align}
     \frac{1}{\frac{1}{A_{22}}\frac{\epsilon}{\epsilon - \gamma} + i\frac{\epsilon^2}{\epsilon - \gamma}p_0} \xrightarrow{\gamma=0} \frac{1}{\frac{1}{A_{22}} + i\epsilon p_0}=a_{\rm F} ,
\end{align}
In general, if $\gamma$ is nonzero, the constant term in the denominator of $f_{11}^{\rm G}$ is different from the correct scattering length $a_{\rm G}$, so an analysis using the Flatt\'{e} amplitude where the scattering length appears in $f_{11}$ may not give the correct scattering length.
\section{Application}
\label{sec-3}
\begin{figure}[tbp]
     \centering
     \includegraphics[width=6cm,clip,bb=0 0 730 500]{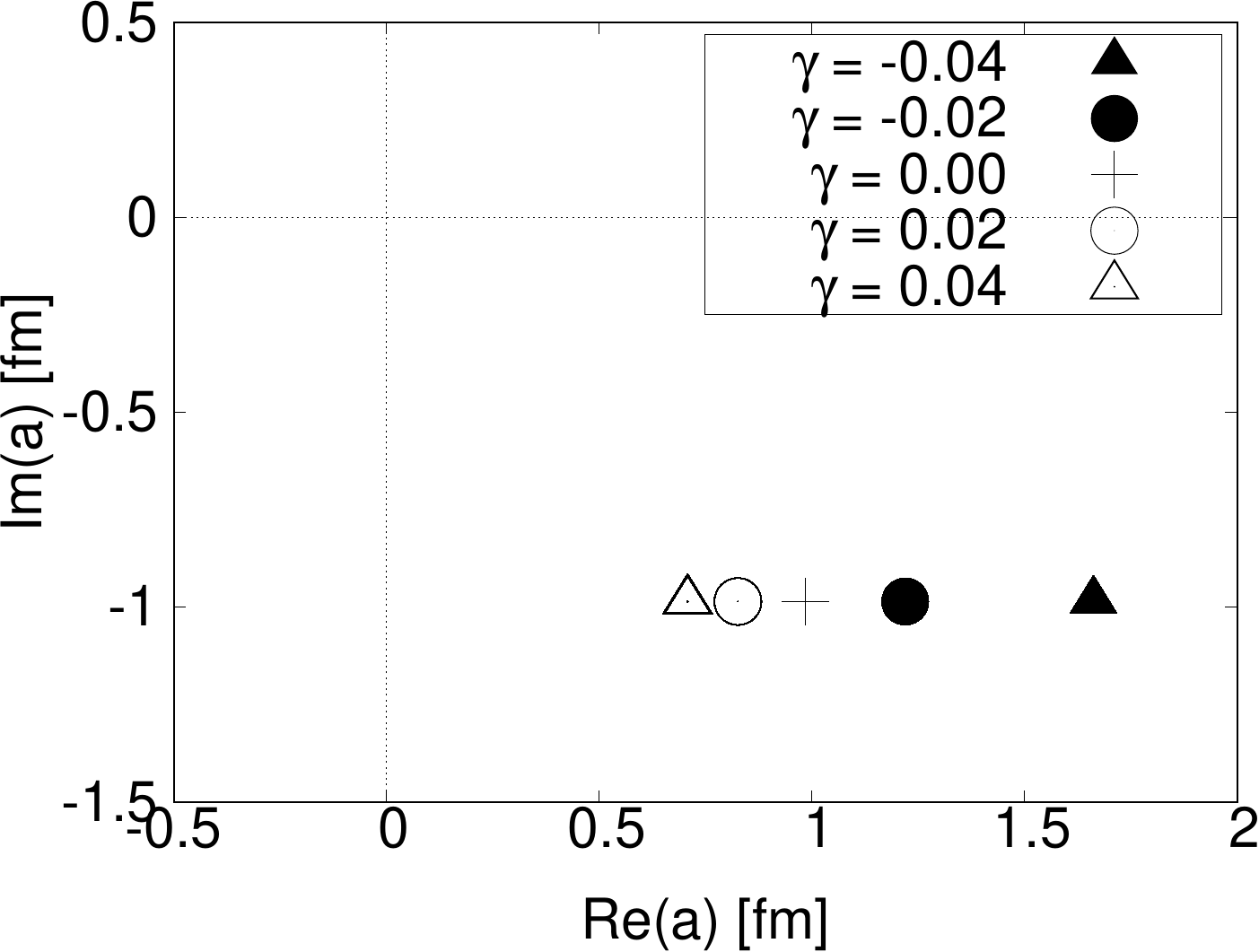}
     \caption{Real and imaginary parts of the scattering length when $\gamma$ is varied from $-0.04$ to $+0.04$. The cross represents the scattering length of the Flatt\'{e} amplitude with $\gamma=0$.}
     \label{fig-a}       
\end{figure}

In order to verify the effect of the value of $\gamma$ on the scattering length $a_{\rm G}$, we fix the constant term of the denominator of $f_{11}^{\rm G}$ and vary $\gamma$. In this study, we consider the $\pi\pi$-$K\bar{K}$ system with $f_{0}(980)$, which has already been analyzed by the Flatt\'{e} amplitude. In Ref.~\cite{CMD-2:1999znb}, the constant term in the denominator of $f_{\pi\pi}$ corresponding to $f^{\rm G}_{11}$ is determined to be $-1.0 - 1.0i$~GeV in the analysis using the Flatt\'{e} amplitude. In this case, two conditions are imposed to the parameters $A_{22}, \gamma, \epsilon$. In order to compare the scattering lengths $a_{\rm G}$ and $a_{\rm F}$ of the Flatt\'{e} amplitude, we calculate the scattering length $a_{\rm G}$ with difference values of $\gamma$. The change of the scattering length $a_{\rm G}$ when $\gamma$ is varied from $-0.04$ to $+0.04$ is shown in Fig.\ref{fig-a}.

In Fig. \ref{fig-a}, the point represented by the cross ($\gamma=0$) corresponds to the scattering length $a_{\rm F}$ of the Flatt\'{e} amplitude, and the general scattering length $a_{\rm G}$ deviates from $a_{\rm F}$ by $\sim 0.5$ fm when $\gamma$ is changed from $-0.04$ to $+0.04$. In the present case, the imaginary part of $a_{\rm G}$ does not depend on $\gamma$ as seen in Fig. \ref{fig-a}. This property can be analytically shown by the imaginary part of Eq.~\eqref{eq: the scattering length of fG}. We find that the scattering length $a_{\rm G}$ varies quantitatively for different $\gamma$. Therefore, the scattering length determined from the Flatt\'{e} amplitude $a_{\rm F}$ with $\gamma=0$ may deviate from the correct scattering length $a_{\rm G}$ with $\gamma \neq 0$ numerically.
\section{Summary}
\label{sec-4}
In this study, we discuss the properties of general scattering amplitudes with the channel couplings. First, the EFT amplitude and the Flatt\'{e} amplitude are compared, showing that the EFT amplitude does not reduce to the Flatt\'{e} amplitude  directly. Next, we solve this problem by introducing a new parametrization of the EFT amplitude to construct the general amplitude that includes both the EFT amplitude and the Flatt\'{e} amplitude. Finally, by applying the general amplitude to the $\pi\pi$-$K\bar{K}$ system and quantitatively comparing the correct scattering length with the one determined from the Flatt\'{e} amplitude, we show that the scattering length of the Flatt\'{e} amplitude may deviate from the correct scattering length by about $0.5$ fm.
%
%

\end{document}